# Evolution of the bulk properties, structure, magnetic order, and superconductivity with Ni doping in CaFe$_{2-x}$Ni$_x$As$_2$

Neeraj Kumar[1], Songxue Chi[2,3], Ying Chen[2,3], Kumari Gaurav Rana[1], A. K. Nigam[1], A. Thamizhavel[1], W. Ratcliff II[2], S. K. Dhar[1], and Jeffrey W. Lynn[2*]

[1]Department of Condensed Matter Physics and Materials Science, Tata Institute of Fundamental Research, Homi Bhabha Road, Colaba, Mumbai 400 005, India.
[2]NIST Center for Neutron Research, National Institute of Standards and Technology, Gaithersburg, Maryland 20899, USA
[3]Department of Materials Science and Engineering, University of Maryland, College Park, Maryland 20742, USA

*Corresponding author

**Abstract**

Magnetization, susceptibility, specific heat, resistivity, neutron and x-ray diffraction have been used to characterize the properties of single crystalline CaFe$_{2-x}$Ni$_x$As$_2$ as a function of Ni doping for x varying from 0 to 0.1. The combined first-order structural and magnetic phase transitions occur together in the undoped system at 172 K, with a small decrease in the area of the *a-b* plane along with an abrupt increase in the length of the *c*-axis in the orthorhombic phase. With increasing x the ordered moment and transition temperature decrease, but the transition remains sharp at modest doping while the area of the *a-b* plane quickly decreases and then saturates. Warming and cooling data in the resistivity and neutron diffraction indicate hysteresis of ≈2 K. At larger doping the transition is more rounded, and decreases to zero for x ≈ 0.06. The susceptibility is anisotropic for all values of x. Electrical resistivity for x = 0.053 and 0.06 shows a superconducting transition with an onset of nearly 15 K which is further corroborated by substantial diamagnetic susceptibility. For the fully superconducting sample there is no long range magnetic order and the structure remains tetragonal at all temperature, but there is an anomalous *increase* in the area of the *a-b* plane in going to low T. Heat capacity data show that the density of states at the Fermi level increases for x ≥ 0.053 as inferred from the value of Sommerfeld coefficient γ. The regime of superconductivity is quite restrictive, with a maximum T$_C$ of 15 K and an upper critical field H$_{c2}$=14 T. Superconductivity disappears in the overdoped region.

## I. Introduction

The discovery of superconductivity at elevated temperatures in Fe-As based pnictides[1,2] has led to an intense research activity on these fascinating materials. With regard to the 122 class of materials, the parent compounds have the composition AFe$_2$As$_2$ (A = Ca, Ba, Sr and Eu), crystallizing in the well known tetragonal ThCr$_2$Si$_2$ type structure. The Fe-sublattice undergoes a spin density wave (SDW) transition in the temperature range 100-200 K depending upon A and there is a concomitant structural transition from tetragonal to orthorhombic symmetry[3-7]. Hole-doping at the A-site induced by dopants like Na$^+$ and K$^+$ and electron doping achieved by replacing Fe partially with 3d elements like Co and Ni suppress the SDW and the structural transitions, leading to superconductivity with high T$_C$[8-11]. More recently, it has been reported that superconductivity can also be stabilized by substituting 4d and 5d elements like Rh, Pd and Ir for Fe in AFe$_2$As$_2$ compounds[12-13]. T$_C$ as high as 38 K in hole-doped Na and K systems and ~24 K in electron-doped systems has been achieved. There have also been reports of doping on the As site and EuFe$_2$As$_{1.4}$P$_{0.6}$ has been shown to be a superconductor with a T$_C$ of 26 K[14]. The SDW transition is severely affected by doping and the precise interrelationship between the residual magnetic correlations and superconductivity is not yet fully understood.

We have reported previously that a single crystal of $CaFe_{1.94}Co_{0.06}As_2$ shows superconducting behavior with an onset $T_C$ of 17 K[15]. The signatures of the SDW transition at $T_{SDW}$ = 171 K in the parent $CaFe_2As_2$ as seen in the resistivity and heat capacity data are absent in the superconducting Co-doped compound. $^{57}Fe$ Mössbauer spectra in the doped compound show a quadrupole-split spectrum at 300 K. Below 90 K, a broadening in the wings is observed which progressively increases with the decrease of temperature. The latter may arise either due to a residual magnetic hyperfine interaction or due to relaxation effects. In the present work, we have carried out a detailed study of Ni substitution at the Fe site in $CaFe_2As_2$ by making a series of single crystals of nominal composition $CaFe_{2-x}Ni_xAs_2$ ( x = 0, 0.05, 0.1, 0.15, 0.2, 0.3, 0.4, 0.5, 0.6, 0.8 and 1) by the high temperature Sn-flux method. The properties of single crystals were probed using the techniques of powder x-ray diffraction, single crystal neutron diffraction, magnetization, electrical resistivity and heat capacity. We find that superconductivity is induced in a narrow region of Ni doping in $CaFe_{2-x}Ni_xAs_2$ with an onset $T_C$ of ~ 15 K. A $T_C$ maximum of 20.5 K and 9.8 K has been reported in recent studies on $BaFe_{2-x}Ni_xAs_2$[11] and $SrFe_{2-x}Ni_xAs_2$ single crystals[16], respectively. We mention here that the actual compositions of our single crystals inferred from electron probe microanalysis (EPMA) give a Ni content which is typically an order of magnitude less than the nominal composition, as seen below.

## II. Experimental Procedures

Single crystals of $CaFe_{2-x}Ni_xAs_2$ for various x were grown by the high temperature solution growth Sn flux method using the same heating and cooling protocol as mentioned in Ref. 15. Laue patterns were recorded to find the crystal symmetry and orientation. Powder x-ray diffraction spectra of all the samples were taken to determine the phase purity and the lattice constant values at 300 K by crushing a few pieces of single crystals, using Cu Kα radiation (PANanalytical X'pert Pro). Electron probe micro-analysis (EPMA) was done on all the crystals to determine their composition accurately using a CAMECA SX-100. Magnetization and heat capacity were measured in a SQUID magnetometer and PPMS (Quantum Design), respectively. Electrical resistivity was measured on a home built, automated set-up.

Neutron diffraction measurements were carried out to study the structural and magnetic phase transitions below room temperature. Data were collected on the BT-7 and BT-9 triple-axis spectrometers at the NIST Center for Neutron Research. The neutron wavelength employed was 2.359 Å using a pyrolytic graphite (PG) (002) monochromator and analyzer (when employed), and a PG filter sufficiently thick to completely suppress higher-order reflections to achieve a monochromatic incident beam. Söller collimations were varied from 60′-50′-S-50′-100′ full-width-at-half maximum (FWHM) for coarse resolution conditions, to 10′-10′-S-10′-10′ FWHM for tight resolution conditions to detect the orthorhombic splitting below the structural phase transition. We denote positions in momentum space using $Q = (H, K, L)$ in reciprocal lattice units in which Q (in Å$^{-1}$) = ($2\pi H /a$, $2\pi K /b$, $2\pi L /c$). The samples were mounted on an aluminum plate with Al foil to avoid any significant stresses, as even modest pressure or stress can dramatically affect the properties.[17,18] They were aligned in the [H, 0, L] zone inside a sealed aluminum container with helium exchange gas and mounted on the cold finger of a closed cycle helium refrigerator. Some data were also taken in a top-loading helium cryostat. For the neutron data, error bars where indicated are statistical in origin and represent one standard deviation.

## III. Results and Discussion

EPMA analysis was performed on very well polished single crystalline samples both on the cross-section and as well on the top-surface of the as-grown single crystal. The EPMA data averaged over ten to fifteen different spots showed that the actual Ni content in the crystals is about one tenth of the starting composition. The nominal and EPMA derived Ni compositions are listed in Table I. Henceforth, we will refer to the single crystals by their actual Ni composition. We believe the Ni solubility is determined by the highest temperature to which the contents are heated during the process of crystal growth. The "un-reacted Ni" is presumably mostly embedded inside the Sn flux and then removed with the flux, or it may partially replace Fe in $CaFe_4As_3$ which is also formed in small quantity during crystal growth. The latter can be easily scrapped off the surface of 122 crystals and does not give rise to any contamination problem. The powder x-ray diffraction patterns of the crushed single crystals showed the peaks corresponding to the 122 phase. For some samples a few weak intensity peaks were also detected which did not correspond to known peaks for Sn. The powder patterns thus show that the single crystals are predominantly single phase and the amount of Sn incorporated into the lattice, if any, is below the detection limit of x-ray (< 2 %). Figure 1 shows the variation of the tetragonal lattice constants $a$ and $c$ with respect to x. Overall the lattice parameter $a$ increases while $c$ and the unit cell volume decrease with the substitution of Ni for Fe. Similar behavior has been reported for $SrFe_{2-x}M_xAs_2$ (M = Rh, Ir and Pd)[13], $BaFe_{2-x}Ru_xAs_2$[19] and $BaFe_{2-x}Ni_xAs_2$[20].

A. Electrical Resistivity

Figure 2 shows the electrical resistivity of $CaFe_{2-x}Ni_xAs_2$ alloys for various x. At very low doping the resistivity of Ni-doped crystals is similar to the parent $CaFe_2As_2$. The SDW and associated structural transition are characterized by a relatively sharp upturn in the resistivity. Using this indicator, we find that increasing the Ni concentration leads to lower transition temperatures. The onset of the transition, shown by an upward arrow in the figure, decreases from $T_{SDW}$ = 171 K in $CaFe_2As_2$ to 150 K in $CaFe_{1.985}Ni_{0.015}As_2$. The transition is first order exhibiting a hysteresis of ~ 2 K in the cooling and heating data recorded for x = 0.006, 0.008 and 0.015, as shown in Fig. 2 for x = 0.015. The drop in the resistivity below 3.5 K is due to the presence of traces of Sn on the surface of single crystals. With further increase of Ni concentration the upturn due to the SDW transition becomes broader while shifting further to lower temperatures. A drop in the resistivity occurs at ~15 K for x = 0.027 and 0.030, which at higher Ni doping (x = 0.053 and 0.06) develops into a full superconducting transition with an onset $T_C$ of 15 K (determined by the same method as used in Ref. 15) and a transition width of nearly 3 K for both values of x. The electrical resistivity decreases monotonically with temperature down to the onset of the superconducting transition in $CaFe_{1.94}Ni_{0.06}As_2$, apparently showing no signature of the SDW transition. However, for x = 0.053 there is a change in the slope of the resistivity plot near 80 K which is related to residual magnetic order at this composition, as discussed below when the neutron data are presented. At higher Ni concentrations of x = 0.075 and 0.1 superconductivity vanishes showing that superconductivity is induced in a narrow range of Ni-doping. For $CaFe_{1.925}Ni_{0.075}As_2$ a drop in resistivity near 15 K, similar to that observed for x = 0.027 and 0.030, indicates the presence of isolated filamentary superconducting regions in the sample. With further increase in Ni concentration the drop at 15 K disappears.

The expression $\rho = \rho_0 + AT^n$ was found to provide a good fit to the resistivity data of $CaFe_{1.94}Ni_{0.06}As_2$ between 20 and 50 K with n = 1.9 and the coefficient A = 36x$10^{-3}$ µΩ-cm. For the non-superconducting $CaFe_{1.9}Ni_{0.1}As_2$, the corresponding values are n = 2.04 and A = 5.1x$10^{-3}$ µΩ-cm for a fit between 4 and 40 K as shown in Fig.2(b). The observation of $T^2$ behavior over a fairly extended temperature range indicates the dominance of electron-electron scattering in the system and the formation of a Fermi liquid regime at low temperatures for these two compositions. A $T^2$ behavior of resistivity up to ~68 K has been reported in $BaCo_2As_2$[21] and $Ba(Fe-Ni)_2As_2$[20], both of which are non-superconducting.

The electrical resistivity of $CaFe_{1.94}Ni_{0.06}As_2$ in applied magnetic fields with J // $a$-$b$ plane and B applied along the $c$-axis is shown in Fig. 3(a). $T_C$ decreases and the superconducting transition broadens with increasing B, which is typical for type–II superconductors. The resistivity drop is clearly seen even for fields as high as 9 T, which indicates that the upper critical field in this compound is high, similar to the corresponding Co-doped compound $CaFe_{1.94}Co_{0.06}As_2$ and other Fe-As based superconductors. The temperature dependence of the upper critical field is shown in Fig. 3(b). The upper critical field $H_{c2}$ varies almost linearly with field and does not show any kind of saturation for fields as high as 9 T. The slope $dH_{c2}/dT_C$ is estimated to be -1.3 T/K. Using the single-band Werthamer-Helfand-Hohenberg formula[22], $H_{c2}(0) = -0.697 T_C (dH_{c2}/dT)_{Tc}$, $H_{c2}(0)$ is estimated to be 14 T for a $T_C$ of 15 K.

B. Magnetic Susceptibility

Figure 4 shows the temperature dependence of the magnetic susceptibility of $CaFe_{2-x}Ni_xAs_2$ in the temperature range from 1.8 to 300 K for some values of x. The data are taken in an applied magnetic field of 1T. The magnitude of the susceptibility at 300 K shows minor variations with Ni-doping. Like the parent $CaFe_2As_2$, the susceptibility of Ni-doped compounds is largely anisotropic. We remark that the upturn in the susceptibility at low temperatures may not be intrinsic. For example, M vs. H plots at 1.8 K measured up to 7 T in $CaFe_{1.985}Ni_{0.015}As_2$ give $\chi = 0.58 \times 10^{-3}$ and $0.53 \times 10^{-3}$ emu/mol Oe (note that 1 emu/mol Oe = $4\pi\ 10^{-6}\ m^3$/mol) for H // a-b and H // c, respectively, which are lower than the corresponding values obtained in M vs. T runs. The signature of the SDW transition at low doping is clearly seen by a sharp drop in the susceptibility along both the main crystallographic directions at temperatures which are in good agreement with resistivity data.

A diamagnetic transition with an onset of 13.5 K is seen in both $CaFe_{1.947}Ni_{0.053}As_2$ and $CaFe_{1.94}Ni_{0.06}As_2$ in an applied field of 0.005 T (Fig. 4), but the magnitude of the signal is larger in the latter compound. An approximate estimate of the superconducting fraction may be obtained by taking the demagnetization factor, N, for H // a-b to be zero while $1/(1-N) \approx 1 + w/2d$ for H // c, where w is the geometric mean of the two lateral dimensions of our flat platelet and 2d is its thickness[23]. Such a procedure gives a superconducting fraction of nearly 75 and 100 % for H // a-b and H //c in $CaFe_{1.94}Ni_{0.06}As_2$, respectively. In $CaFe_{1.947}Ni_{0.053}As_2$ the fractions thus calculated are 11 and 4.5 %. While these numerical estimates need not be overemphasized, qualitatively they show that a slight increase of Ni concentration increases the superconducting fraction substantially and that a Ni concentration of x = 0.06 is most likely close to the optimum doping for superconductivity in $CaFe_{2-x}Ni_xAs_2$.

It may also be noted that barring the low temperature upturn, the susceptibility decreases with decreasing temperature in all the samples. In local moment systems the susceptibility shows a Curie - Weiss like increase with decreasing temperature and even in itinerant systems the susceptibility typically decreases with increasing temperature. However, here the susceptibility increases linearly with T above $T_{SDW}$ and above ~ 100 K for x = 0.06. A linear-T dependence above $T_{SDW}$ (~ 136 K) up to 700 K has been reported in $BaFe_2As_2$ for both H // a-b and H //c[24]. Theoretically it has been argued[25] that the linear-T susceptibility above $T_{SDW}$ arises from strong antiferromagnetic correlations in a regime characterized by antiferromagnetic fluctuations with local spin density wave correlations. Such a picture implies that antiferromagnetic fluctuations are present even in the superconducting $CaFe_{1.94}Ni_{0.06}As_2$. The susceptibility of our samples at 300 K is comparable to that of a $BaFe_2As_2$ single crystal grown by the self-flux method[25] and therefore indicates a similar magnitude of antiferromagnetic correlations.

C. Heat Capacity

Figure 5 shows the heat capacity of compounds between 1.8 and 200 K. Peaks in the heat capacity data for x = 0, 0.006, 0.008 and 0.02 at ~ 172, 162, 157.5 and 145 K, correlate well with the anomalies seen in the corresponding resistivity and magnetization data. Barely discernible anomalies, which become broader with increasing x, are seen centered at 113 and 82 K for x = 0.03 and 0.053 respectively. An expanded view of the heat capacity plots around the peak region (bottom inset Fig. 5) for x = 0.006 and x = 0.02 suggests a possible splitting of the anomaly. Recent work on $BaFe_{2-x}Co_xAs_2$[20,26] reports that the single magnetic/structural phase transition observed in undoped $BaFe_2As_2$ is split into two distinct phase transitions on Co-doping, inferred from the observation of two distinct features in the heat capacity and resistivity data and further supported by $d(\chi T)/dT$ data. This has been speculated to be either due to the separation of structural and magnetic phase transitions or to a distribution in the Co concentration. Our resistivity data plotted in Fig. 1 do not show any splitting into two distinct regimes, and this is confirmed by the neutron diffraction data discussed below. Although we see an apparent splitting of the heat capacity peak at the SDW transition for particular compositions (0.006 and 0.020), based on the overall evidence from neutron diffraction, heat capacity and electrical resistivity, we believe that the magnetic and structural transitions occur together when doping $CaFe_2As_2$ with Ni.

The low temperature heat capacity data are plotted in the form of $C/T$ vs. $T^2$ in the upper inset of Fig. 5. A least-squares fit of the expression $C/T = \gamma + \beta T^2$ (where $\gamma$ and $\beta$ have their usual meaning) to the data below ~7 K yields $\gamma$ = 7.2, 6.2, 25.8, 22.4, 25.7 and 24.4 mJ/mol $K^2$ for x = 0, 0.008, 0.027, 0.053, 0.06 and 0.1, respectively. The values of the Debye temperature $\theta_D$, obtained from the corresponding values of $\beta$ are 241, 255, 269, 234, 219 and 251 K, respectively. For pure $CaFe_2As_2$, the present value of $\gamma$ is intermediate between 4.7[27] and 8.2 mJ/mol $K^2$ (Ref. 28). At present we do not understand this relatively large sample dependent variation in the measured values of $\gamma$. For x $\geq$ 0.027, the data show that $\gamma$ increases significantly suggesting an appreciable enhancement in the density of states at the Fermi level. For superconducting $CaFe_{1.94}Ni_{0.06}As_2$ we do not observe a jump in the specific heat at $T_C$ nor does $C/T$ decrease exponentially below $T_C$. The apparent lack of anomaly at $T_C$ may be due to the relatively large transition width (3 K) of the superconducting transition.

### D. Neutron Diffraction

The magnetic and structural phase transitions were investigated in detail using neutron diffraction. Below the structural phase transition the tetragonal symmetry is distorted into orthorhombic symmetry, and is clearly indicated by the splitting of the tetragonal [2,2,0] Bragg peak into [4,0,0] and [0,4,0] orthorhombic peaks.[29] An example of the observed splitting just below the transition (150 K) is shown in Fig. 6(b), compared to above the transition at 190 K (Fig. 6(a)), for the lightly doped single crystal $CaFe_{1.994}Ni_{0.006}As_2$. The scattering associated with both the structural and magnetic phase transitions was measured in detail as a function of temperature for most of the compositions (Table 1), and an example of a map of the scattering is shown in Fig. 7 for the same composition of $CaFe_{1.994}Ni_{0.006}As_2$. Upon cooling, we note that the structural peak abruptly splits into two peaks (Fig. 7(b)).

One of the interesting features to note in Fig. 7(a,b) is that the orthorhombic splitting is not symmetric about the tetragonal Bragg peak, in contrast to what is observed in the $SrFe_2As_2$ system[5]. This means that there is a change in the area of the *a-b* plane when the orthorhombic distortion occurs, with the area being reduced in the orthorhombic phase. The temperature dependence of the [0,0,4] peak is shown in Fig. 7(d), which indicates that this reduction in the area of the *a-b* plane is compensated to some extent by an increase in the length of the *c*-axis. This also contrasts with the $SrFe_2As_2$ system, where no change in the *c*-axis could be detected[30]. For the $CaFe_2As_2$ system, both the undoped material and with Ni doping up to the region where superconductivity develops, there is an abrupt change in the *c*-axis lattice parameter at the structural transition. We also note that there is a small thermal hysteresis of a few degrees K in the lattice parameters (Fig. 7(a-d)), which is mirrored in the magnetic order parameter as shown in Fig. 7(e, f). This is in good agreement with the hysteresis observed in the resistivity data. We will discuss the magnetic order in more detail below.

Figure 8(a-d) compares the nature of the orthorhombic distortion as a function of doping. For two compositions well below the superconducting regime (Fig. 8(a, b) we see that the size of the distortion is about the same, but the transition is not as abrupt for the higher doping. For x=0.053 where superconductivity just appears, the distortion is difficult to discern clearly but its size is much reduced in value. For the higher doping of x=0.060 where robust superconductivity is observed, no structural distortion is detected with the current (high) resolution. Nevertheless, there is a strong temperature dependence of the tetragonal lattice parameter. Interestingly, this is in the opposite direction to that at lower x, which exhibit an orthorhombic splitting. There the *a-b* plane area decreases in going into the orthorhombic phase, while here the area expands as we cool to low temperature. It will be interesting to see if band calculations can reproduce this dependence on doping. The size of the splitting as a function of doping is shown in Fig. 9(a), where we see that the splitting is essentially constant as a function of doping until superconductivity appears, and then quickly drops to zero.

As noted above, the orthorhombically split peaks are not positioned symmetrically about the tetragonal peak, and this means that the area of the *a-b* plane of the unit cell changes at the transition. We define the asymmetry α(x) by

$$\alpha(x) = 1 - \left|\frac{a' - a}{b - a}\right| \quad , \quad (1)$$

where *a'* indicates the *a* axis (in orthorhombic notation) in the orthorhombic phase. For the undoped material the asymmetry is small in value but clear, while it increases quickly with increasing doping and seems to saturate at higher doping as shown in Fig. 9(b). The data in Fig. 9(b) would suggest that the splitting would become exactly symmetric if a very small amount of 'negative" Ni concentration could be introduced.

The lattice parameters and unit cell volumes are shown in Fig. 10 for the seven compositions where the structural distortion is observed, together with the fully superconducting composition where no structural transition or magnetic order is observed. It may be noted that above the structural phase transition the system is tetragonal so that $a = b$ (as in Fig. 1), while in the orthorhombic phase the unit cell is rotated by 45° with $a_{OR} \approx b_{OR} \approx \sqrt{2}\, a_T$. For the undoped system there is an obvious jump in all three lattice parameters, which is accompanied by a strong anomaly in the unit cell volume. For the x=0.006 doping the size of the changes in the lattice parameters is approximately the same, but the abrupt anomaly at the transition is gone, and the overall change in the unit cell volume is smaller. With increasing x we see that the change in the volume of the unit cell decreases. For x=0.053, where superconductivity first appears, we find only a subtle structural transition. This is indicated in Fig. 8(c), where we compare the intensity and position of the [4,0,0] and [0,4,0] peaks for four values of x. We see that the small splitting of the structural peak is actually just a broadening/shoulder and this makes it difficult to accurately determine where the onset of the transition occurs. At the highest concentration we have investigated (x=0.06), we can no longer detect any structural anomaly.

We remark that the asymmetry reported here is based on Bragg reflections that are directly related to the distortion in the *a-b* plane. Likewise, the change in the *c*-axis lattice parameter is determined directly by measuring *c*-axis reflections. We note that the present results are qualitatively different than reported by Ni, *et al.*[27] in their Fig. 8, where they indicate a huge asymmetry where both orthorhombic peaks jump on the same side of the tetragonal peak, and the jump in the *c*-axis corresponds to a decrease in *c* in the orthorhombic phase. However, for those data the *a* and *b* lattice parameters were extracted from a single reflection with mixed indices and this does not provide accurate values for *a, b* and *c* in the orthorhombic phase.

Turning now to the magnetic ordering, for the undoped system we find that the magnetic ordering occurs at the same temperature as the structural transition, with the intensity of the magnetic Bragg peaks appearing abruptly, in good agreement with previous work[3]. Figure 11 shows the [1,0,3] magnetic Bragg intensity for several of the doped systems. At the lower doping levels we see the same type of abrupt behavior. Therefore the magnetic and structural transitions are clearly first-order in nature and occur together. This contrasts with the behavior of the 1111 systems where the magnetic transition is second order in nature and occurs at a lower temperature than the structural transition,[29,31] or Co doped $BaFe_2As_2$ where the transitions separate with doping.[32] With increasing nickel content both the antiferromagnetic ordering temperature and the size of the ordered magnetic moment at low temperature both decrease, as given in Table 2. At larger x the transition appears to become more continuous or smeared, while the magnetic structure continues to be identical to the parent compound. For the x=0.053 composition where some superconductivity first appears, the magnetic order parameter may reflect some distribution of transition temperatures that may be due to a small composition variation which smears the transition. The small superconducting fraction, low SDW temperature and small ordered moment point to macroscopic rather than microscopic coexistence of the magnetic order and superconductivity. The onset of the structural transition is also difficult to quantify in the neutron data, while the resistivity indicates a transition of ≈80 K which is the same as for the magnetic order. Given the weak scattering and possibly broadened transitions we cannot rule out that the two transitions might occur at different temperatures, and could be continuous similar to what is observed in undoped $BaFe_2As_2$([7,33]). For the fully superconducting sample, on the other hand, we find no evidence for any long range magnetic order (or structural distortion). Indeed the trend of the data indicate that both $T_N$ and the ordered moment vanish at or very near the onset of bulk superconductivity, as has been found in the $CeFeAsO_{1-x}F_x$ system[31]. It will be interesting to see if the $T_C$ vs. Fe-As-Fe tetrahedral angle in this Ni doped system obeys the same relation as has been found in other systems[17,31], but this will require full and detailed structural refinements to be carried out (on powders) as a function of x. A summary of the structural and magnet transition temperatures (on warming) determined by neutron diffraction are given in Table 2, together with the ordered magnetic moment in the ground state.

From the electrical resistivity, magnetic susceptibility and the neutron data we have constructed the magnetic phase diagram of $CaFe_{2-x}Ni_xAs_2$ for various values of x as shown in Fig.12. The structural/magnetic transition temperature monotonically decreases with the increase in x, and a transition to the zero resistance state at $T_C \sim 15$ K is observed only for x = 0.053 and 0.060. On the other hand, a weakly diamagnetic response below ~ 6 K is seen for x = 0.027 and 0.075 but the resistivity shows only a drop at 15 K which does not, however, develop into a fully superconducting state at lower temperatures. It is quite likely that bulk superconductivity extends for x values beyond 0.06, but to establish the upper limit additional samples with x slightly greater than 0.06 would need to be fabricated, characterized, and investigated. Similarly, additional compositions between 0.053 and 0.060 would be needed to further characterize the phase boundaries and order parameters in this regime.

Lastly, we compare the temperature-composition (T-x) phase diagram obtained in the present work on $CaFe_{2-x}Ni_xAs_2$ with the one's reported for $AFe_{2-x}T_xAs_2$ (A = Ba and Sr; T = Ni, Co, Rh, Pd and Ir). The suppression of the SDW transition with increasing x is generic to all these systems. The disappearance of the SDW transition in the Ba and Sr analogs is simultaneously accompanied by the onset of the superconducting transition, with $T_C$ showing a dome-like dependence on the doping concentration x. In $CaFe_{2-x}Ni_xAs_2$ single crystals superconductivity is induced in a limited doping range of x which is relatively smaller. Whether this small range is related to the tetrahedral angle or points to basic differences between the Ba/Sr and Ca based systems remains to be explored.

## IV. Conclusions

In summary, we have grown single crystals of $CaFe_{2-x}Ni_xAs_2$ (x = 0 to 0.1) by high temperature solution growth using Sn as a flux. The electrical resistivity, magnetic susceptibility, heat capacity and neutron diffraction measurements clearly establish that the SDW and the structural transition are first order in nature and occur at the same temperature. From the intensity maps of the structural peak, it is found that the orthorhombic splitting is asymmetric about the tetragonal Bragg peak indicating that there is a decrease in the area of the *a-b* plane when the orthorhombic distortion occurs. As the Ni concentration is increased the $T_{SDW}$ temperature and ordered moment decrease, and vanish when superconductivity develops. However, superconductivity occurs over a very narrow region of the dopant concentration. In contrast to the decrease in the *a-b* plane area at the tetragonal-orthorhombic transition at smaller x, in the superconducting regime the structure remains tetragonal but the area increases in going to low T. The optimal dopant concentration for superconductivity is found to be x = 0.06 with a $T_C$ of 15 K. Similar to the other isostructural superconducting compounds, the $H_{c2}$ is high in this compound, close to 14 T at base temperature.

Acknowledgments
We thank Andreas Kreyssig for helpful discussions.

Table 1. Nominal starting composition and the actual crystal composition in $CaFe_{2-x}Ni_xAs_2$ as determined by EPMA.

| CaFe$_{2-x}$Ni$_x$As$_2$ | |
|---|---|
| Nominal composition (x) | Actual composition (x) |
| 0 | 0 |
| 0.05 | 0.006 |
| 0.10 | 0.008 |
| 0.15 | 0.015 |
| 0.20 | 0.020 |
| 0.30 | 0.027 |
| 0.40 | 0.030 |
| 0.50 | 0.053 |
| 0.60 | 0.060 |
| 0.80 | 0.075 |
| 1.0 | 0.1 |

Table II. Structural and magnetic transition temperatures obtained from the neutron diffraction measurements. Values are determined on warming. The ordered antiferromagnetic moment is determined at low temperatures, and decreases with increasing x.

| Ni doping | Structural Transition (K) | $T_N$ | Ordered Moment ($\mu_B$) |
|---|---|---|---|
| 0 | 172(1) | 172(1) | 0.80(5)[3] |
| 0.006 | 166(1) | 168(1) | 0.65(5) |
| 0.008 | 161(1) | 161(1) | 0.62(5) |
| 0.015 | 151(1) | 151(1) | 0.48(5) |
| 0.020 | 146(1) | 148(2) | 0.16(3) |
| 0.027 | 129(1) | 128(2) | 0.06(2) |
| 0.053 | ≈80 | ≈80 | 0.04(1) |
| 0.060 | No transition | No transition | No moment |

Figure 1. (color online) Variation of the tetragonal lattice constants *a* and *c* with Ni concentration at 300 K.

Figure 2. (color online) (a) Temperature dependence of normalized electrical resistivity of $CaFe_{2-x}Ni_xAs_2$ for various values of x for J // ab-plane. The inset in the plot for x = 0.015 indicates the hysteresis observed in the resistivity while cooling and warming. The upward arrow indicates the SDW transition and the superconducting transition is indicated by the downward arrow. For lower concentrations of x, the drop in the resistivity at low temperature is due to the superconducting transition of Sn (trace amounts of residual flux). The solid black line for x = 0.06 and 0.10 indicates the $T^2$ behavior of the resistivity. (b) Low temperature part of the resistivity data for x = 0.06 and 0.10. The solid line indicates the fit to the Fermi liquid relation (see text for details). The drop at 3.7 K is the superconductivity of trace quantities of Sn present in the sample.

Figure 3. (color online) Temperature dependence of the electrical resistivity in $CaFe_{1.94}Ni_{0.06}As_2$ for current parallel to ab plane and the field parallel to c-axis for various applied fields. The temperature dependence of the upper critical field of $CaFe_{1.94}Ni_{0.06}As_2$. The solid line in (b) is estimated based on the WHH theory as mentioned in the text.

Figure 4 (color online) Temperature dependence of magnetic susceptibility of $CaFe_{2-x}Ni_xAs_2$ for various concentrations of x. The upward arrow indicates the SDW transition and the downward arrow indicates the superconducting transition. The open circle symbols represent field parallel to ab-plane and the filled circle symbols represent field parallel to c-axis.

Figure 5 (color online) Temperature dependence of heat capacity in the range 1.8 to 200 K for various concentrations of x in $CaFe_{2-x}Ni_xAs_2$. The lower inset shows the enlarged region around the peak for x = 0.006 and 0.020. The upper inset shows the C/T versus $T^2$ behavior.

Figure 6. (color online )Example of diffraction scans of the (a) tetragonal (2,2,0) peak above the structural transition and (b) the orthorhombically split (4,0,0) and (0,4,0) peaks below the structural transition, for a single crystal of $CaFe_{1.994}Ni_{0.006}As_2$. The solid curves are least-squares fits to the (Gaussian) instrumental resolution.

Figure 7. (color online ) Intensity maps of the structural and magnetic Bragg peaks through the phase transition for $CaFe_{1.994}Ni_{0.006}As_2$ on warming and cooling, where we observe a few degrees hysteresis. (a, b) (4,0,0) and (0,4,0) structural Bragg peaks (orthorhombic notation) on warming and cooling, respectively. Above the transition these peaks are equivalent. Note that below the phase transition the peaks are not symmetric about the tetragonal position, which means that there is a change in the area of the *a-b* plane, in particular a decrease in the area in the orthorhombic phase. (c,d) (0,0,4) *c*-axis Bragg peak on warming and cooling, respectively. Note that there is a sudden decrease in the position of the peak, which translates into an increase in the *c*-axis lattice parameter. This tends to compensate for the decrease in the area of the *a-b* plane. (e,f) sudden disappearance on warming and appearance on cooling of the (1,0.3) magnetic Bragg peak, showing that the transition is a combined structural and magnetic one.

Figure 8. (color online ) Temperature maps of the $(2,2,0)_T$ to $(4,0,0)_O$, $(0,4,0)_O$ peaks for four different compositions. At lower x the transition is very sharp, but it lowers in temperature and broadens with increasing x. For x=0.053 where the superconducting state just appears, the scattering just broadens in the orthorhombic phase. For x=0.063 where robust superconductivity occurs no structural distortion is detected, while the lattice parameter is strongly temperature dependent.

Figure 9. (color online ) (a) Orthorhombic splitting as a function of composition. Little change in the splitting is observed until superconductivity appears, where it quickly drops to zero. (b). Asymmetry of the splitting, defined by Eq. (1), versus composition. For the undoped compound the asymmetry is small but obvious, and quickly grows and saturates with doping until the transition abruptly disappears. Solid curves are a guide to the eye.

Figure 10. (color online ) (a-g) Temperature dependence of the lattice parameters and volume of the unit cell for the seven compositions where the structural distortion is detected. Here orthorhombic notation is used in the tetragonal phase to facilitate direct comparisons. (h) Temperature dependence of the lattice parameters and width of the (4,0,0) Bragg peak for the fully superconducting sample, which remains tetragonal.

Figure 11. (color online ) Temperature dependence of (a-d) the (1,0,3) magnetic Bragg peak, which has the strongest intensity, for four different dopings. At small x there is a jump in the magnetic scattering indicating that the transition is first-order and occurs together with the structural phase transition (as shown in Fig. 7). At higher x (d) the ordered moment is smaller and the transition appears to be closer to continuous, which could mean that it is smeared due to some compositional inhomogeneity, or it is intrinsically second order. (e,f) scattering for the (1,0,3) and (1,0,1) magnetic Bragg peaks for x=0.053. The (1,0,3) is still the strongest peak so that no change in the magnetic structure is indicated.

Figure 12. (color online) The phase diagram determined from the resistivity, magnetic susceptibility and neutron diffraction measurements of $CaFe_{2-x}Ni_xAs_2$ with respect to various values of x. The superconductivity region is almost flat and the optimum doping concentration of Ni lies close to x= 0.060 as shown by the green colored region. The red shaded region on the left shows the (perhaps macroscopic) coexistence of SDW order and filamentary superconductivity on the left, but the bulk resistivity does not attain zero value. The right hand side red-hatched region indicates a resistivity drop near 15 K but the sample does not become superconducting.

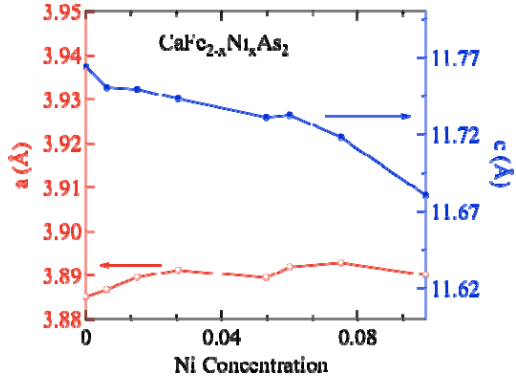

Fig. 1

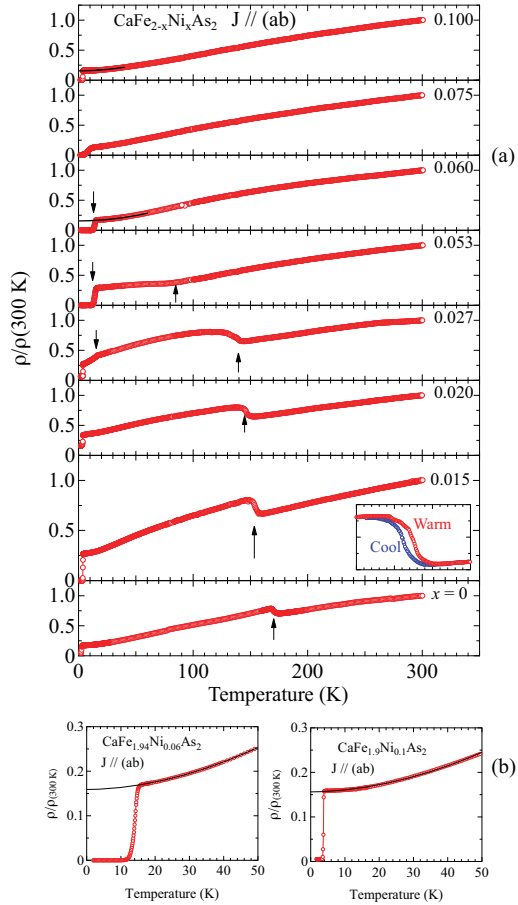

Fig. 2.

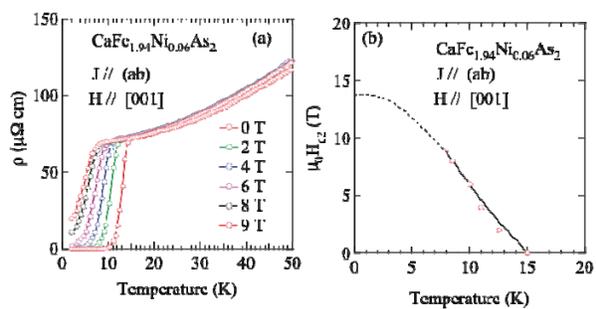

Fig. 3.

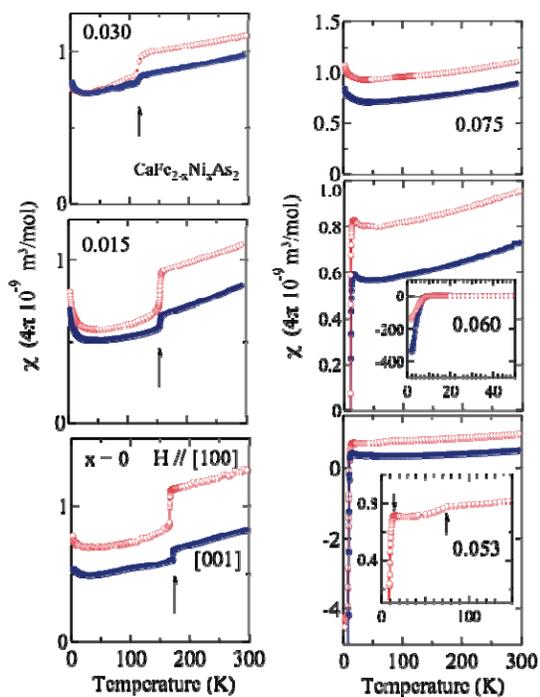

Figure 4

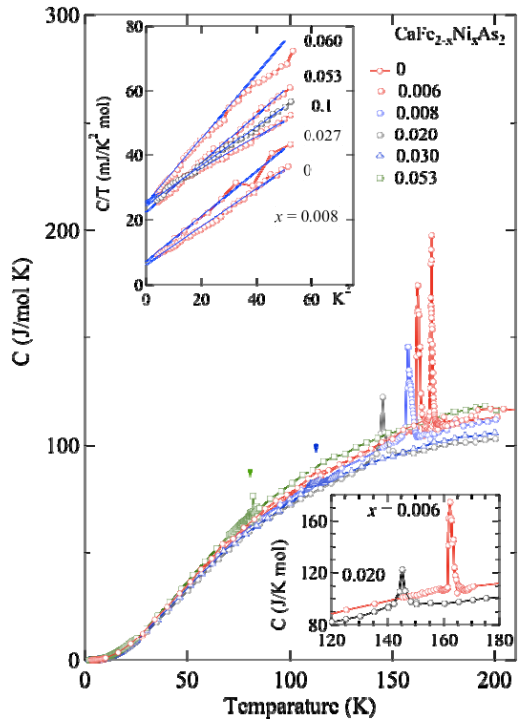

Figure 5

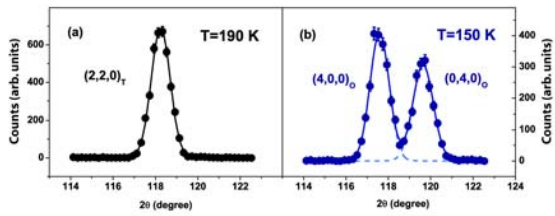

Figure 6.

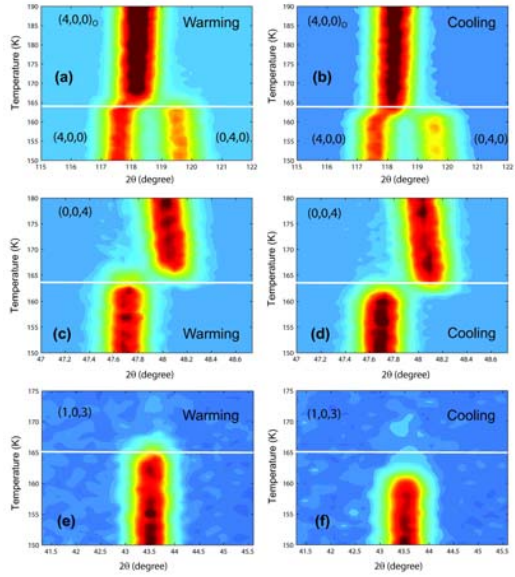

Figure 7.

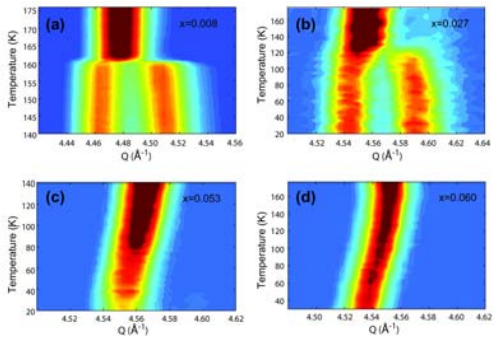

Figure 8.

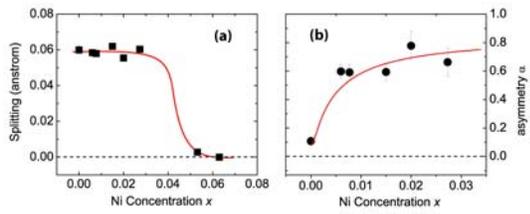

Figure 9.

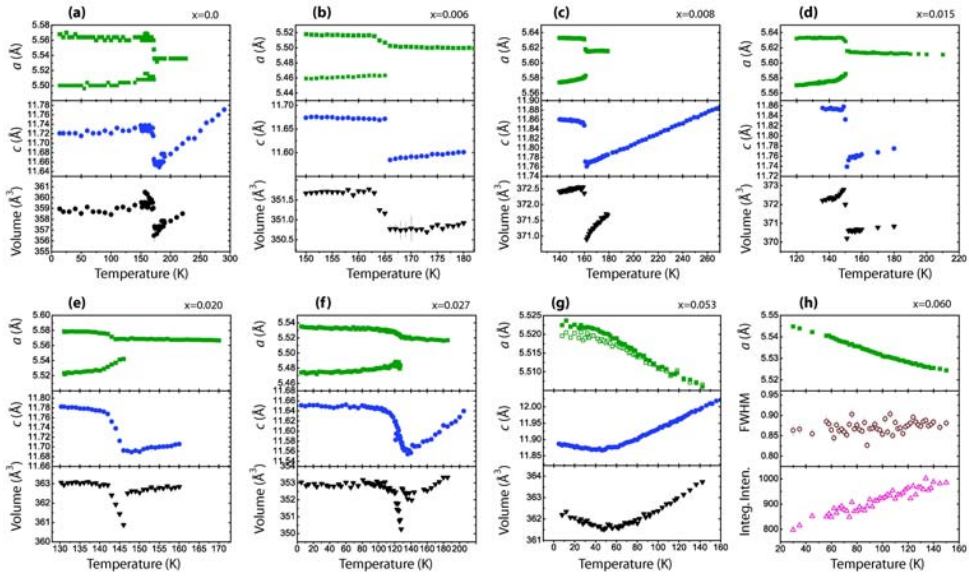

Figure 10.

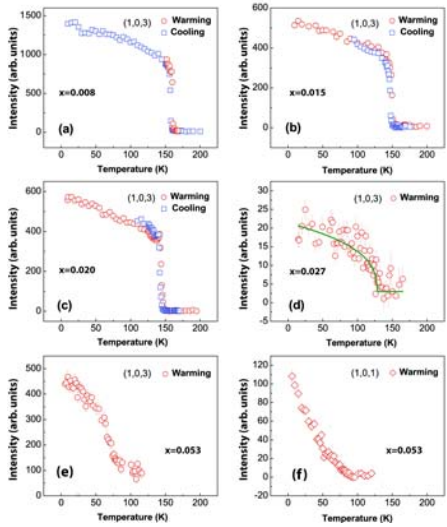

Fig. 11

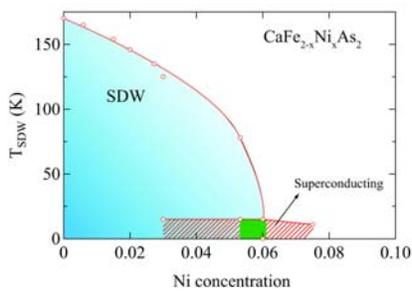

Fig. 12.

16